\input{aipcheck}

\documentclass[
    ,final            
  ,numberedheadings 
  ]
  {aipproc}

\layoutstyle{6x9}
\def\Journal#1#2#3#4{{#1} {\bf #2}, #3 (#4)}

\def\RNC{\em Rivista Nuovo Cimento}

\def\NIMA{{\em Nucl. Instrum. Methods} A}

\def\PLB{{\em Phys. Lett.}  B}
\def\PRL{\em Phys. Rev. Lett.}
\def\PRD{{\em Phys. Rev.} D}

\def\GaC{\em Gravitation and Cosmology}

\def\JETPL{\em JETP Lett.}

\def\CQG{\em Class. Quantum Grav.}
\def\APJ{\em Astrophys. J.}
\def\SCI{\em Science}
\def\MPLA{{\em Mod. Phys. Lett.}  A}
\def\IJTP{\em Int. J. Theor. Phys.}
\def\NJP{\em New J. of Phys.}
\def\JHEP{\em JHEP}
\def\BWP{\em Bled Workshops in Physics}

\def\s{{\,\rm s}}
\def\g{{\,\rm g}}
\def\eV{\,{\rm eV}}
\def\keV{\,{\rm keV}}
\def\MeV{\,{\rm MeV}}
\def\GeV{\,{\rm GeV}}
\def\TeV{\,{\rm TeV}}
\def\sv{\left<\sigma v\right>}
\def\({\left(}
\def\){\right)}
\def\cm{{\,\rm cm}}

\def\kpc{{\,\rm kpc}}
\def\beq{\begin{equation}}
\def\eeq{\end{equation}}
\def\bea{\begin{eqnarray}}
\def\eea{\end{eqnarray}}

\begin{document}

\title{Dark Atoms of Dark Matter from New Stable Quarks and Leptons }


\author{Maxim Yu. Khlopov}{
  address={APC laboratory 10, rue Alice Domon et L\'eonie Duquet, 75205
Paris Cedex 13, France}
  ,altaddress={Centre for Cosmoparticle Physics "Cosmion",
    National Research Nuclear University "Moscow Engineering Physics Institute", 115409 Moscow, Russia}
}

\begin{abstract}
 The nonbaryonic dark matter of the Universe can consist of
new stable charged leptons and quarks, if they are
hidden in elusive "dark atoms" of composite dark matter. Such possibility can be compatible with the severe constraints on anomalous isotopes, if there exist stable particles with charge  -2  and there are no stable particles  with charges +1 and -1. These conditions cannot be realized in supersymmetric models, but can be
satisfied in several recently developed alternative scenarios.
The excessive -2 charged particles are bound with
primordial helium in O-helium "atoms", maintaining specific
nuclear-interacting form of the Warmer than Cold Dark Matter. The puzzles of
direct dark matter searches appear in this case as a reflection
of nontrivial nuclear physics of O-helium.
\end{abstract}

\maketitle


\section{Introduction}

  According to the modern cosmology, the dark matter, corresponding to
$25\%$ of the total cosmological density, is nonbaryonic and
consists of new stable particles. Such particles (see e.g.
\cite{book,Cosmoarcheology,Bled07} for review and reference) should
be stable, provide the measured dark matter density and be decoupled
from plasma and radiation at least before the beginning of matter
dominated stage. The easiest way to satisfy these conditions is to
involve neutral elementary weakly interacting massive particles
(WIMPs). SUSY Models provide a list of possible WIMP candidates:
neutralino, axino, gravitino etc.

However it may not be the only
particle physics solution for the dark matter problem.

One of such alternative solutions is based on the existence of heavy stable charged particles bound in neutral "dark atoms". This idea of composite dark matter was first proposed by S.L. Glashow in \cite{Glashow}.
According to \cite{Glashow} stable tera-U-quarks with electric charge +2/3
forms stable $(UUU)$ +2 charged "clusters", which in combination with two -1
charged stable tera-electrons E produce neutral $[(UUU)EE]$ tera-helium "atoms" that
behave like WIMPs. The main problem for this solution is the over-abundance of positively charged species
bound with
ordinary electrons, which behave as anomalous isotopes of hydrogen
or helium. This problem turned to be unresolvable, because in the early Universe
as soon as primordial helium is formed it would
capture all the free $E^-$ and form positively charged $(He E)^+$ ion,
preventing any further suppression of positively charged species \cite{Fargion:2005xz}.
Therefore, in order to avoid anomalous isotopes overproduction,
stable particles with charge -1 (and corresponding antiparticles)
should be absent, so that stable negatively charged particles should
have charge of -2 only \cite{Khlopov:2006dk}.
Stable particles with the charge of -2 turned to be the only solution that saved the idea of dark atoms of dark matter.

One should mention here that stable double charged particles can hardly find place in SUSY models, but there
exist several alternative elementary particle frames, in which heavy
stable -2  charged species, $X^{--}$, are predicted:
\begin{itemize}
\item[(a)] AC-leptons, predicted
in the extension of standard model, based on the approach
of almost-commutative geometry \cite{Khlopov:2006dk,5,FKS,bookAC}.
\item[(b)] Technileptons and
anti-technibaryons in the framework of walking technicolor
models (WTC) \cite{KK,Sannino:2004qp}.
\item[(c)] stable "heavy quark clusters" $\bar U \bar U \bar U$ formed by anti-$U$ quark of fourth
 \cite{Khlopov:2006dk,Q,I,lom} or 5th
 \cite{Norma} generation.
\end{itemize}
All these models also
predict corresponding +2 charge particles. If these positively charged particles remain free in the early Universe,
they can recombine with ordinary electrons in anomalous helium, which is strongly constrained in the
terrestrial matter. Therefore cosmological scenario should provide mechanism of suppression of anomalous helium.
There are two possibilities, which require two different mechanisms of such suppression:
\begin{itemize}
\item[(i)] The abundance of anomalous helium in the Galaxy may be significant, but in the terrestrial matter
there exists a recombination mechanism suppressing this abundance below experimental upper limits.
\item[(ii)] Free positively charged particles are already suppressed in the early Universe and the abundance
of anomalous helium in the Galaxy is negligible.
\end{itemize}
These two possibilities correspond to to two different cosmological scenarios of dark atoms. The first is
realized in the scenario with AC leptons, which form WIMP-like neutral AC atoms.

The second, which is considered here following \cite{KK,KK2,unesco,iwara,Levels,DMDA,PSTJ} assumes charge asymmetric case with the excess of $X^{--}$ that form atom-like states with
primordial helium. After it is formed
in the Big Bang Nucleosynthesis, $^4He$ screens the
$X^{--}$ charged particles in composite $(^4He^{++}X^{--})$ {\it
O-helium} ``atoms'' \cite{I}. All the
 For different models of $X^{--}$ these "atoms" are also
called ANO-helium \cite{Khlopov:2006dk,lom}, Ole-helium
\cite{Khlopov:2006dk,FKS} or techni-O-helium \cite{KK}. We'll call
them all O-helium (or denote by $OHe$) \cite{Levels,I2} in our further discussion of their cosmological evolution, following the guidelines of \cite{I2}.

In all these forms of O-helium, $X^{--}$ behaves either as lepton or
as specific "heavy quark cluster" with strongly suppressed hadronic
interaction. Therefore O-helium interaction with matter is
determined by nuclear interaction of its helium shell. These neutral primordial
nuclear interacting objects contribute to the modern dark matter
density and play the role of a nontrivial form of strongly
interacting dark matter \cite{Starkman,McGuire:2001qj}.

If new stable species belong to non-trivial
representations of electroweak SU(2) group, sphaleron transitions at
high temperatures provide the relationship between baryon asymmetry
and excess of -2 charge stable species. It makes possible to relate the density of asymmetric O-helium dark matter with the baryon density.

Here after a brief review of main features of OHe evolution in the Universe we
concentrate on the nuclear radiative capture of O-helium in underground detectors that might explain puzzles of dark matter searches.
\section{\label{OHeformation} Formation and evolution of O-helium in the Universe}
\subsection{O-helium dark matter}
Following \cite{Khlopov:2006dk,KK,I,lom,unesco,iwara,Levels,PSTJ,I2} consider
charge asymmetric case, when excess of $X^{--}$ provides effective
suppression of positively charged species.

In the period $100\s \le t \le 300\s$  at $100 \keV\ge T \ge T_o=
I_{o}/27 \approx 60 \keV$, $^4He$ has already been formed in the
SBBN and virtually all free $X^{--}$ are trapped by $^4He$ in
O-helium ``atoms" $(^4He^{++} X^{--})$. Here the O-helium ionization
potential is\footnote{The account for charge distribution in $He$
nucleus leads to smaller value $I_o \approx 1.3 \MeV$
\cite{Pospelov}.} \beq I_{o} = Z_{x}^2 Z_{He}^2 \alpha^2 m_{He}/2
\approx 1.6 \MeV,\label{IO}\eeq where $\alpha$ is the fine structure
constant,$Z_{He}= 2$ and $Z_{x}= 2$ stands for the absolute value of
electric charge of $X^{--}$.  The size of these ``atoms" is
\cite{FKS,I} \beq R_{o} \sim 1/(Z_{x} Z_{He}\alpha m_{He}) \approx 2
\cdot 10^{-13} \cm \label{REHe} \eeq Here and further, if not
specified otherwise, we use the system of units $\hbar=c=k=1$.

The analysis \cite{Levels} favors
Bohr-atom-like structure of O-helium, assumed in
\cite{Khlopov:2006dk,KK,I,lom,unesco,iwara,I2}. However, the size of
He, rotating around $X^{--}$ in this Bohr atom, turns out to be of
the order and even a bit larger than the radius $r_o$ of its Bohr
orbit, and the corresponding correction to the binding energy due to
non-point-like charge distribution in He is significant.

Bohr atom like structure of OHe seems to provide a possibility to
use the results of atomic physics for description of OHe interaction
with matter. However, the situation is much more complicated. OHe
atom is similar to the hydrogen, in which electron is hundreds times
heavier, than proton, so that it is proton shell that surrounds
"electron nucleus". Nuclei that interact with such "hydrogen" would
interact first with strongly interacting "proton" shell and such
interaction can hardly be treated in the framework of perturbation
theory. Moreover in the description of OHe interaction the account
for the finite size of He, which is even larger than the radius of
Bohr orbit, is important. One should consider, therefore, the
analysis, presented in \cite{Khlopov:2006dk,KK,I,lom,unesco,iwara,Levels,I2,Khlopov:2008rp}, as only a first step approaching true nuclear physics of OHe.

Due to nuclear interactions of its helium shell with nuclei in
the cosmic plasma, the O-helium gas is in thermal equilibrium with
plasma and radiation on the Radiation Dominance (RD) stage, while
the energy and momentum transfer from plasma is effective. The
radiation pressure acting on the plasma is then transferred to
density fluctuations of the O-helium gas and transforms them in
acoustic waves at scales up to the size of the horizon.

At temperature $T < T_{od} \approx 200 S^{2/3}_3\eV$ the energy and
momentum transfer from baryons to O-helium is not effective
\cite{KK,I} because $$n_B \sv (m_p/m_o) t < 1,$$ where $m_o$ is the
mass of the $OHe$ atom and $S_3= m_o/(1 \TeV)$. Here \beq \sigma
\approx \sigma_{o} \sim \pi R_{o}^2 \approx
10^{-25}\cm^2\label{sigOHe}, \eeq and $v = \sqrt{2T/m_p}$ is the
baryon thermal velocity. Then O-helium gas decouples from plasma. It
starts to dominate in the Universe after $t \sim 10^{12}\s$  at $T
\le T_{RM} \approx 1 \eV$ and O-helium ``atoms" play the main
dynamical role in the development of gravitational instability,
triggering the large scale structure formation. The composite nature
of O-helium determines the specifics of the corresponding dark
matter scenario.

At $T > T_{RM}$ the total mass of the $OHe$ gas with density $\rho_d
= (T_{RM}/T) \rho_{tot} $ is equal to
$$M=\frac{4 \pi}{3} \rho_d t^3 = \frac{4 \pi}{3} \frac{T_{RM}}{T} m_{Pl}
(\frac{m_{Pl}}{T})^2$$ within the cosmological horizon $l_h=t$. In
the period of decoupling $T = T_{od}$, this mass  depends strongly
on the O-helium mass $S_3$ and is given by \cite{KK}\beq M_{od} =
\frac{T_{RM}}{T_{od}} m_{Pl} (\frac{m_{Pl}}{T_{od}})^2 \approx 2
\cdot 10^{44} S^{-2}_3 \g = 10^{11} S^{-2}_3 M_{\odot}, \label{MEPm}
\eeq where $M_{\odot}$ is the solar mass. O-helium is formed only at
$T_{o}$ and its total mass within the cosmological horizon in the
period of its creation is $M_{o}=M_{od}(T_{od}/T_{o})^3 = 10^{37}
\g$.

On the RD stage before decoupling, the Jeans length $\lambda_J$ of
the $OHe$ gas was restricted from below by the propagation of sound
waves in plasma with a relativistic equation of state
$p=\epsilon/3$, being of the order of the cosmological horizon and
equal to $\lambda_J = l_h/\sqrt{3} = t/\sqrt{3}.$ After decoupling
at $T = T_{od}$, it falls down to $\lambda_J \sim v_o t,$ where $v_o
= \sqrt{2T_{od}/m_o}.$ Though after decoupling the Jeans mass in the
$OHe$ gas correspondingly falls down
$$M_J \sim v_o^3 M_{od}\sim 3 \cdot 10^{-14}M_{od},$$ one should
expect a strong suppression of fluctuations on scales $M<M_o$, as
well as adiabatic damping of sound waves in the RD plasma for scales
$M_o<M<M_{od}$. It can provide some suppression of small scale
structure in the considered model for all reasonable masses of
O-helium. The significance of this suppression and its effect on the
structure formation needs a special study in detailed numerical
simulations. In any case, it can not be as strong as the free
streaming suppression in ordinary Warm Dark Matter (WDM) scenarios,
but one can expect that qualitatively we deal with Warmer Than Cold
Dark Matter model.

Being decoupled from baryon matter, the $OHe$ gas does not follow
the formation of baryon astrophysical objects (stars, planets,
molecular clouds...) and forms dark matter halos of galaxies. It can
be easily seen that O-helium gas is collisionless for its number
density, saturating galactic dark matter. Taking the average density
of baryon matter one can also find that the Galaxy as a whole is
transparent for O-helium in spite of its nuclear interaction. Only
individual baryon objects like stars and planets are opaque for
it.

O-helium atoms can be destroyed in astrophysical processes, giving
rise to acceleration of free $X^{--}$ in the Galaxy. If the mechanisms of $X^{--}$ acceleration are effective, the
anomalous low $Z/A$ component of $-2$ charged $X^{--}$ can be
present in cosmic rays at the level $X/p \sim n_{X}/n_g \sim
10^{-9}S_3^{-1},$ and be within the reach for PAMELA and AMS02
cosmic ray experiments  \cite{KK2,Mayorov}.

Collisions
in the galactic bulge can lead to excitation of O-helium. If 2S
level is excited, pair production dominates over two-photon channel
and positron production is sufficient to explain the
excess in positron annihilation line from bulge\cite{Finkbeiner:2007kk}, measured by
INTEGRAL \cite{integral}.

It should be noted that the nuclear cross section of the O-helium
interaction with matter escapes the severe constraints
on strongly interacting dark matter particles
(SIMPs) \cite{Starkman,McGuire:2001qj} imposed by the XQC experiment
\cite{XQC}. Therefore, a special strategy of direct O-helium  search
is needed, as it was proposed in \cite{Belotsky:2006fa}.

\section{O-helium effects in underground detectors}
\subsection{O-helium in the terrestrial matter} The evident
consequence of the O-helium dark matter is its inevitable presence
in the terrestrial matter, which appears opaque to O-helium and
stores all its in-falling flux.

After they fall down terrestrial surface, the in-falling $OHe$
particles are effectively slowed down due to elastic collisions with
matter. Then they drift, sinking down towards the center of the
Earth with velocity \beq V = \frac{g}{n \sigma v} \approx 80 S_3
A^{1/2} \cm/\s. \label{dif}\eeq Here $A \sim 30$ is the average
atomic weight in terrestrial surface matter, $n=2.4 \cdot 10^{24}/A \cm^{-3}$
is the number density of terrestrial atomic nuclei, $\sigma v$ is the rate
of nuclear collisions, $m_o \approx M_X+4m_p=S_3 \TeV$ is the mass of O-helium,
$M_X$ is the mass of the $X^{--}$ component of O-helium, $m_p$ is the mass of proton and $g=980~ \cm/\s^2$.

Near the Earth's surface, the O-helium abundance is determined by
the equilibrium between the in-falling and down-drifting fluxes.

The in-falling O-helium flux from dark matter halo is
$$
  F=\frac{n_{0}}{8\pi}\cdot |\overline{V_{h}}+\overline{V_{E}}|,
$$
where $V_{h}$-speed of Solar System (220 km/s), $V_{E}$-speed of
Earth (29.5 km/s) and $n_{0}=3 \cdot 10^{-4} S_3^{-1} \cm^{-3}$ is the
local density of O-helium dark matter. For qualitative estimation we
don't take into account here velocity dispersion and distribution of particles
in the incoming flux that can lead to significant effect.

At a depth $L$ below the Earth's surface, the drift timescale is
$t_{dr} \sim L/V$, where $V \sim 400 S_3 \cm/\s$ is given by
Eq.~(\ref{dif}). It means that the change of the incoming flux,
caused by the motion of the Earth along its orbit, should lead at
the depth $L \sim 10^5 \cm$ to the corresponding change in the
equilibrium underground concentration of $OHe$ on the timescale
$t_{dr} \approx 2.5 \cdot 10^2 S_3^{-1}\s$.

The equilibrium concentration, which is established in the matter of
underground detectors at this timescale, is given by
\begin{equation}
    n_{oE}=\frac{2\pi \cdot F}{V} = n_{0}\frac{n \sigma v}{4g} \cdot
    |\overline{V_{h}}+\overline{V_{E}}|,
\end{equation}
where, with account for $V_{h} > V_{E}$, relative velocity can be
expressed as
$$
    |\overline{V_{o}}|=\sqrt{(\overline{V_{h}}+\overline{V_{E}})^{2}}=\sqrt{V_{h}^2+V_{E}^2+V_{h}V_{E}sin(\theta)} \simeq
$$
$$
\simeq V_{h}\sqrt{1+\frac{V_{E}}{V_{h}}sin(\theta)}\sim
V_{h}(1+\frac{1}{2}\frac{V_{E}}{V_{h}}sin(\theta)).
$$
Here $\theta=\omega (t-t_0)$ with $\omega = 2\pi/T$, $T=1yr$ and
$t_0$ is the phase. Then the concentration takes the form
\begin{equation}
    n_{oE}=n_{oE}^{(1)}+n_{oE}^{(2)}\cdot sin(\omega (t-t_0))
    \label{noE}
\end{equation}

So, if OHe reacts with nuclei in the matter of underground detector, there are two parts of the signal
of such reaction: constant and annual
modulation, as it is expected in the strategy of dark matter search
in DAMA experiments \cite{Bernabei:2003za,Bernabei:2008yi}.

\subsection{O-helium interaction with nuclei}

The explanation \cite{unesco,iwara,Levels,DMDA} of the results of
DAMA/NaI \cite{Bernabei:2003za} and DAMA/LIBRA
\cite{Bernabei:2008yi} experiments is based on the idea that OHe,
slowed down in the matter of detector, can form a few keV bound
state with nuclei, in which OHe is situated \textbf{beyond} the
nucleus. Therefore the positive result of these experiments is
explained by reaction
\begin{equation}
A+(^4He^{++}X^{--}) \rightarrow [A(^4He^{++}X^{--})]+\gamma
\label{HeEAZ}
\end{equation}
with nuclei in DAMA detector.

The approach of \cite{unesco,iwara,Levels,PSTJ} assumes the following
picture:
\begin{itemize}
\item[(i)] At the distances larger, than its size, OHe is neutral,
being only the source of a Coulomb field of $X^{--}$ screened by
$He$ shell. Owing to the negative sign of $Z_{X}=-2$, this
potential provides attraction of nucleus to OHe.
\item[(ii)] Then helium shell of OHe starts to feel Yukawa exponential tail of
attraction of nucleus to $He$ due to scalar-isoscalar nuclear
potential.
\item[(iii)] Nuclear attraction results in the polarization of
OHe and the mutual attraction of nucleus and OHe is changed by
Coulomb repulsion of $He$ shell.
\item[(iv)] When helium is completely merged with the nucleus the interaction is
reduced to the oscillatory potential of $X^{--}$ with
homogeneously charged merged nucleus with the charge $Z+2$.
\end{itemize}
It should be noted that scalar-isoscalar nature of He
nucleus excludes its nuclear interaction due to $\pi$ or $\rho$
meson exchange, so that the main role in its nuclear interaction
outside the nucleus plays $\sigma$ meson exchange, on which nuclear
physics data are not very definite. The nuclear potential depends on
the mass $\mu$ of the $\sigma$-meson, coupling to nucleon $g^2$ and on the relative distance between He and nucleus. It would imply axial symmetric quantum mechanical description. In the approximation of spherical symmetry nuclear attraction beyond the nucleus was taken into account in \cite{Levels}
in a two different ways:
\begin{itemize}
\item[(m)] The nuclear Yukawa potential was averaged over the
orbit of He in OHe,
\item[(b)] The potential was taken at the position of He most close to the nucleus.
\end{itemize}
These two cases (m) and (b) correspond to the larger and smaller
distance effects of nuclear force, respectively, so that the true
picture should be between these two extremes.

\begin{figure}
            \includegraphics[height=.3\textheight]{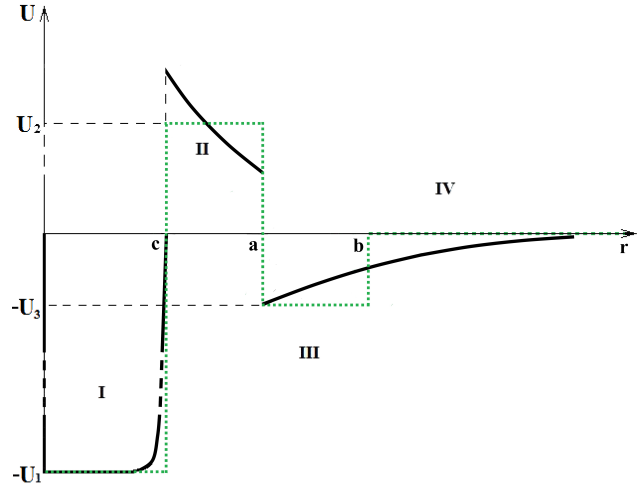}
        \caption{The approximation of rectangular well for potential of OHe-nucleus system.}\label{pic1}
    \end{figure}

To simplify the solution of Schrodinger equation the
rectangular potential was considered in \cite{Levels}
that consists of
\begin{itemize}
\item[(i)] a potential well with the depth $U_1$ at $r<c=R$,
where $R$ is the radius of nucleus;
\item[(ii)] a rectangular dipole Coulomb
potential barrier $U_2$ at $R \le r<a=R+R_o+r_{he}$, where $r_{he}$
is radius of helium nucleus;
\item[(iii)] an outer potential well $U_3$,
formed by the Yukawa nuclear interaction and
residual Coulomb interaction.
\end{itemize}
 It lead to the approximate potential, presented on Fig. \ref{pic1}.

Solutions of Schrodinger equation for each of the four regions,
indicated on Fig. \ref{pic1}, are given in textbooks (see
e.g.\cite{LL3}) and their sewing determines the condition, under
which a low-energy  OHe-nucleus bound state appears in the region
III.
\subsection{Low energy bound state of O-helium with nuclei}
\begin{figure}
            \includegraphics[height=.4\textheight]{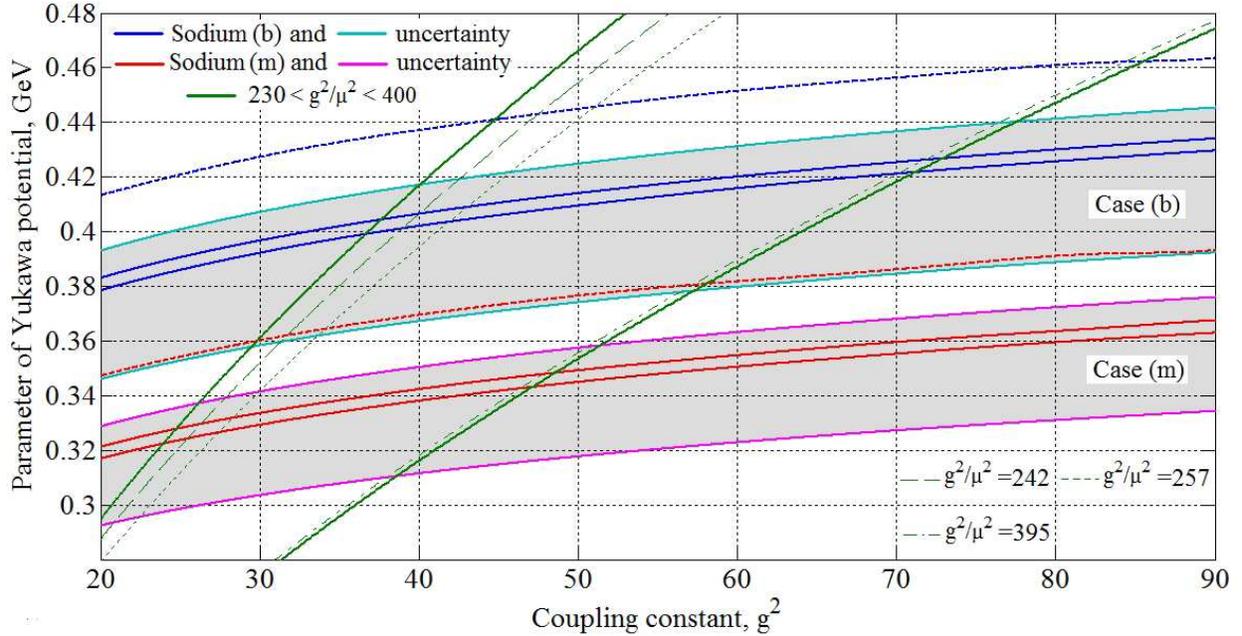}
        \caption{The regions of parameters $\mu$ and $g^2$, for which Na-OHe system has a level in the interval 4 keV for the cases (m) and (b).
        Two lines determine at $d_o=1.2/(200 \MeV)$ the region of parameters, at which the bound
        system of this element with OHe has a 4 keV level.
        In the region between the two strips the energy of level is below 4 keV.
                There are also indicated the range of $g^2/\mu^2$ (dashed lines) as well as their preferred values
        (thin lines) determined from parametrization of the relativistic ($\sigma-\omega$)
		model for nuclear matter. The uncertainty in the determination of parameter $1.15/(200 \MeV)<d_o <1.3/(200 \MeV)$ results in the
        uncertainty of $\mu$ and $g^2$ shown by the shaded regions surrounding the lines.}\label{Na}
    \end{figure}

The energy of this bound state and its existence strongly depend on
the parameters $\mu$ and $g^2$ of nuclear potential.
On the Fig. \ref{Na} the regions of these parameters, giving 4 keV
energy level in OHe bound state with sodium are presented. Radiative
capture to this level can explain results of DAMA/NaI and DAMA/LIBRA
experiments with the account for their energy resolution
\cite{DAMAlibra}. The lower shaded region on Fig. \ref{Na}
corresponds to the case (m) of nuclear Yukawa potential
averaged over the orbit of He in OHe, while the upper region
corresponds to the case (b) of this potential taken at
the position of He most close to the nucleus. The
result is sensitive to the precise value of $d_o$, which
determines the size of nuclei $R=d_o A^{1/3}$. The range of $g^2/\mu^2$ and their preferred values were determined in \cite{nuclear}. In the calculations \cite{Levels} the mass of OHe was
taken equal to $m_o=1 TeV$, however the results weakly depend on the
value of $m_o>1 TeV$.

It is interesting that the values of $\mu$ on Fig. \ref{Na} are compatible with the results of recent experimental
measurements of mass of sigma meson \cite{sigma}.

The important qualitative feature of this solution is the
restricted range of intermediate nuclei, in which the OHe-nucleus
bound state is possible. The range of nuclei with bound states with OHe
corresponds to the part of periodic table between B and Ti. The
results are very sensitive to the numerical factors of calculations
and the existence of OHe-Ge and OHe-Ga bound states at a narrow
window of parameters $\mu$ and $g^2$ turns to be strongly dependent
on these factors so that change in numbers smaller than 1\% can give
qualitatively  different result for Ge and Ga.
\subsection{Radiative capture of O-helium}

In the essence, the explanation \cite{unesco,iwara,Levels,DMDA} of the results of experiments DAMA/NaI and DAMA/LIBRA is based on the idea that OHe concentration in the matter of DAMA detectors possess annual modulation and its radiative capture
to a few keV bound state with sodium nuclei leads to the corresponding energy release and ionization signal, detected in DAMA experiments.

The rate of radiative capture of OHe by nuclei was calculated
\cite{unesco,iwara,DMDA} with the use of the analogy with the radiative
capture of neutron by proton with the account for: i) absence of M1
transition that follows from conservation of orbital momentum and
ii) suppression of E1 transition in the case of OHe. Since OHe is
isoscalar, isovector E1 transition can take place in OHe-nucleus
system only due to effect of isospin nonconservation, which can be
measured by the factor $f = (m_n-m_p)/m_N \approx 1.4 \cdot
10^{-3}$, corresponding to the difference of mass of neutron,$m_n$,
and proton,$m_p$, relative to the mass of nucleon, $m_N$. In the
result the rate of OHe radiative capture by nucleus with atomic
number $A$ and charge $Z$ to the energy level $E$ in the medium with
temperature $T$ is given by
\begin{equation}
    \sigma v=\frac{f \pi \alpha}{m_p^2} \frac{3}{\sqrt{2}} (\frac{Z}{A})^2 \frac{T}{\sqrt{Am_pE}}.
    \label{radcap}
\end{equation}

Formation of OHe-nucleus bound system leads to energy release of its
binding energy, detected as ionization signal.  In the context of
the approach \cite{unesco,iwara,Levels,DMDA,PSTJ} the existence of annual modulations of
this signal in the range 2-6 keV and absence of such effect at energies above 6 keV
means that binding energy of Na-OHe system in DAMA experiment should
not exceed 6 keV, being in the range 2-4 keV. The amplitude of
annual modulation of ionization signal (measured in counts per day
per kg, cpd/kg) is given by
\begin{equation}
\zeta=\frac{3\pi \alpha \cdot n_o N_A V_E t Q}{640\sqrt{2}
A_{med}^{1/2} (A_I+A_{Na})} \frac{f}{S_3 m_p^2} (\frac{Z_i}{A_i})^2
\frac{T}{\sqrt{A_i m_p E_i}}= a_i\frac{f}{S_3^2} (\frac{Z_i}{A_i})^2
\frac{T}{\sqrt{A_i m_p E_i}}. \label{counts}
\end{equation}
Here $N_A$ is Avogadro number, $i$ denotes Na, for which numerical
factor $a_i=4.3\cdot10^{10}$, $Q=10^3$ (corresponding to 1kg of the
matter of detector), $t=86400 \s$, $E_i$ is the binding energy of
Na-OHe system and $n_{0}=3 \cdot 10^{-4} S_3^{-1} \cm^{-3}$ is the
local density of O-helium dark matter near the Earth. The value of
$\zeta$ should be compared with the integrated over energy bins
signals in DAMA/NaI and DAMA/LIBRA experiments and the result of
these experiments can be reproduced for $E_{Na} = 3 \keV$. The
account for energy resolution in DAMA experiments \cite{DAMAlibra}
can explain the observed energy distribution of the signal from
monochromatic photon (with $E_{Na} = 3 \keV$) emitted in OHe
radiative capture.

At the corresponding values of $\mu$ and $g^2$ there is no binding
of OHe with iodine and thallium \cite{Levels}.

It should be noted that the results of DAMA experiment exhibit also
absence of annual modulations at the energy of MeV-tens MeV. Energy
release in this range should take place, if OHe-nucleus system comes
to the deep level inside the nucleus. This transition implies
tunneling through dipole Coulomb barrier and is suppressed below the
experimental limits.

Effects of OHe-nucleus binding should strongly differ in detectors with the content,
different from NaI. For the chosen range of nuclear parameters, reproducing the results
of DAMA/NaI and DAMA/LIBRA, there are no levels in the OHe-nucleus systems for heavy and very light (e.g. $^3He$) nuclei \cite{Levels}.
In particular, there are no such levels in Xe and most probably in Ge, what causes problems for direct comparison
with DAMA results in CDMS experiment \cite{Akerib:2005kh} and seem to prevent  such comparison in XENON100 \cite{xenon} experiments. Therefore test of results of DAMA/NaI and DAMA/LIBRA
experiments by other experimental groups can become a very difficult
task and the puzzles of dark matter search can reflect the nontrivial properties of composite dark matter.
\section{Conclusions}
The existence of heavy
stable charged particles may not only be compatible with the
experimental constraints but can even lead to composite dark matter
scenario of nuclear interacting Warmer than Cold Dark Matter. This
new form of dark matter can provide explanation of excess of
positron annihilation line radiation, observed by INTEGRAL in the
galactic bulge. The search for stable -2 charge component of cosmic
rays is challenging for PAMELA and AMS02 experiments. Decays of
heavy charged constituents of composite dark matter can provide
explanation for anomalies in spectra of cosmic high energy positrons
and electrons, observed by PAMELA and FERMI. In the context of
our approach search for heavy stable charged quarks and leptons at
LHC acquires the significance of experimental probe for components
of cosmological composite dark matter.

The results of dark matter search in experiments DAMA/NaI and
DAMA/LIBRA can be explained in the framework of our scenario without
contradiction with the results of other groups. Our approach contains distinct features, by which
the present explanation can be distinguished from other recent
approaches to this problem \cite{Edward} (see also for review and
more references in \cite{Gelmini}).

OHe concentration in the matter of underground detectors is
determined by the equilibrium between the incoming cosmic flux of
OHe and diffusion towards the center of Earth. It is rapidly
adjusted and follows the change in this flux with the relaxation
time of few minutes. Therefore the rate of radiative capture of OHe
should experience annual modulations reflected in annual modulations
of the ionization signal from these reactions.
Within the uncertainty of the allowed nuclear
physics parameters there exists a range at which OHe binding energy
with sodium is in the interval 2-4 keV. Radiative capture of OHe to
this bound state leads to the corresponding energy release observed
as an ionization signal in DAMA detector.


An inevitable consequence of the proposed explanation is appearance
in the matter of DAMA/NaI or DAMA/LIBRA detector anomalous
superheavy isotopes of sodium, having the mass roughly by 1 TeV
larger, than ordinary isotopes of this element.

Since in the framework of our approach there should be no OHe radiative capture in detectors, containing heavy or light nuclei,
positive result of experimental search for WIMPs by effect of their
nuclear recoil would be a signature for a multicomponent nature of
dark matter. Such OHe+WIMPs multicomponent dark matter scenarios
naturally follow from AC model \cite{FKS} and can be realized in
models of Walking technicolor \cite{KK2}.

The presented approach sheds new light on the physical nature of dark matter.
In this context positive result of DAMA/NaI and DAMA/LIBRA experiments may be a signature for exciting phenomena of O-helium nuclear physics.
\begin{theacknowledgments}
  I express my gratitude to A.G. Mayorov and E. Yu. Soldatov for collaboration in obtaining the presented results and to Jean-Rene Cudell and A.S.Romanyuk for discussions.
\end{theacknowledgments}






\end{document}